# Electronic anisotropy, magnetic field-temperature phase diagram and their dependence on resistivity in c-axis oriented MgB$_2$ thin films

S. Patnaik*, L.D. Cooley*, A. Gurevich*, A.A. Polyanskii*, J. Jiang*, X.Y. Cai*,
A.A. Squitieri*, M.T. Naus*, M.K. Lee†, J.H. Choi†, L. Belenky†, S.D. Bu†, J. Letteri‡, X. Song*,†, D.G. Schlom‡, S. E. Babcock*,†, C. B. Eom*,† E.E. Hellstrom*,† and D. C. Larbalestier*,†

* Applied Superconductivity Center, University of Wisconsin–Madison, 1500 Engineering Drive, Madison, WI 53706
† Department of Materials Science and Engineering, University of Wisconsin–Madison, 1509 University Avenue, Madison, WI 53706
‡ Department of Materials Science and Engineering, Pennsylvania State University, University Park, PA 16802

(Abstract: April 19, 2001)

**Abstract.** An important predicted, but so far uncharacterized, property of the new superconductor MgB$_2$ is electronic anisotropy arising from its layered crystal structure. Here we report on three c-axis oriented thin films, showing that the upper critical field anisotropy ratio $H_{c2}^{//}/H_{c2}^{\perp}$ is 1.8 to 2.0, the ratio increasing with higher resistivity. Measurements of the magnetic field-temperature phase diagram show that flux pinning disappears at $H^* \approx 0.8 H_{c2}^{\perp}(T)$ in untextured samples. $H_{c2}^{//}(0)$ is strongly enhanced by alloying to 39 T for the highest resistivity film, more than twice that seen in bulk samples.

The discovery of superconductivity at almost 40 K in MgB$_2$ has reawakened the search for high critical temperature $T_c$ in compounds with light elements [1]. In spite of the high $T_c$ of bulk MgB$_2$ samples, the upper critical field $H_{c2}(T)$ at which bulk superconductivity is destroyed and the irreversibility field $H^*(T)$ at which bulk supercurrent densities disappear are both comparatively low. The maximum extrapolations of $\mu_0 H_{c2}(0)$ give 16-18 T, while $H^*(0)$ is about $0.5 H_{c2}(0)$ [2-8]. $\mu_0 H^*(4.2\,K)$ is thus 7 T, well below the 10.5 T irreversibility field of Nb47wt.%Ti, for which $T_c$ is 9 K and $\mu_0 H_{c2}(4.2\,K)$ is ~12 T [9]. At present it is not known whether the low irreversibility field of MgB$_2$ is related to its electronic anisotropy, a problem that is well known in the strongly anisotropic, high-temperature copper-oxide superconductors [10]. Since MgB$_2$ consists of alternating B and Mg sheets, electronic anisotropy has been anticipated [11-14], but its explicit determination has so far been held back by the lack of single crystals. Some hints of the anisotropy have been reported for a hot pressed bulk sample [15] and for separated particles allowed to settle on to a flat surface [16], the anisotropy ratio $\eta = H_{c2}^{//}/H_{c2}^{\perp}$ (i.e. parallel and perpendicular to the Mg and B planes) being reported as 1.1 and 1.73 in these two studies, respectively. Since MgB$_2$ looks promising for applications, the magnitude of its anisotropic properties must be resolved because of their implications for controlling flux pinning, magnetic field, and electronic device limits. Further, anisotropy of the MgB$_2$ crystal structure may also be essential to its high $T_c$, so better understanding of the effects of anisotropy could point to undiscovered nuances of its strong superconductivity and to new compounds with still higher $T_c$.

There has also been rapid progress in making MgB$_2$ thin films, with critical current density $J_c$(4.2 K) values exceeding 1 MA/cm$^2$ [17-19]. In [18], $J_c$ of textured films reached 1-3 MA/cm$^2$ at 4.2 K in a 1 T perpendicular field, the high current densities being attributed to the very fine grain size (10-20 nm) of the MgB$_2$ and to similarly sized MgO particles. In this report, two of the films in [18], together with a new, third film, were used to determine the magnetic anisotropy of MgB$_2$ and the influence of the normal-state resistivity on the properties. We show that $H_{c2}(T)$ is anisotropic with $\eta$ = 1.8 to 2.0, the ratio increasing with increasing normal-state resistivity $\rho$. $H_{c2}^{//}(0)$ is estimated at 39 T for the film with the highest resistivity, well above the 30 T $H_{c2}(0)$ value for Nb$_3$Sn [20]. We also found only weak dependence of this anisotropy on resistivity and $H_{c2}$, thus leading us to believe that we are measuring the intrinsic electronic anisotropy of MgB$_2$. An additional issue is the influence of this anisotropy on thermal fluctuations and flux pinning. Our analysis indicates that thermal fluctuations, while weaker than in copper-oxide superconductors with much higher anisotropies, noticeably suppress $H^*$ below $H_{c2}$. For instance, fluctuations reduce $H^{*\perp}(T)$, the practical limit for untextured, round-wire applications, to about 80% of $H_{c2}^{\perp}(T)$. Thus, although there is no evidence of the need to texture MgB$_2$ in order to avoid grain boundary weak links [2-4], the $J_c$ of untextured forms is limited by the lower values of $H^*$ and $H_{c2}$ perpendicular to the Mg and B planes, limiting applications to ~15 T at 4.2 K on the basis of the best present properties.

MgB$_2$ films were deposited on (111) oriented SrTiO$_3$ single crystal substrates by pulsed laser deposition at room temperature followed by post annealing in Mg vapor at different temperatures, as discussed elsewhere [18]. Of the 2 films from [18], film 1 exhibits an expanded c-axis lattice parameter and higher oxygen preparation condition than film 2, making us believe that film 1 is more heavily alloyed with oxygen relative to film 2. Film 3 was annealed like film 2, but at higher temperature (950°C for 15 min). The film thickness was ~500 nm in each case, although interfacial reaction with the SrTiO$_3$ substrate may reduce this somewhat. Careful selected area diffraction electron microscopy and x-ray diffractometry indicated alignment of the MgB$_2$ c-axis with the substrate normal, the electron microscopy showing also that there was random in-plane alignment on length scales of ~ 1 µm [18]. The full width at half-maximum of the (002) MgB$_2$ rocking curves are 8°-10° for all three films.

Film resistance was measured using the four-probe method at a dc current of 1 mA, applied perpendicular to the magnetic field. Resistance was measured in fixed fields applied parallel or perpendicular to the film plane (i.e. the Mg and B planes) while sweeping the temperature. The resistivity at 40 K for films 1, 2, and 3 is 360, 40, and 38 µΩ·cm, respectively. The zero-field resistive transition curves, shown in Fig. 1, indicate progressively higher critical temperatures for films 1, 2, and 3, the zero-resistance values being 30, 31 and 36 K respectively. $T_c$ is highest for film 3 with lowest resistivity and highest resistivity ratio, $RR = \rho(298) / \rho(40) \sim 2.1$, and lowest for film 1 with highest resistivity and $RR \sim 1$ [18]. In contrast to bulk MgB$_2$ samples, where almost all studies have shown $T_c$ values of ~39 K, thin-film $T_c$ values vary considerably over the range ~24-39 K [17-19, 21-26]. It is plausible that this wide range of $T_c$ values is a consequence of alloying of or disorder produced in the MgB$_2$ during growth. As seen from Fig. 1, the zero-field transition of film 2 is less sharp than that of film 1 or film 3, perhaps pointing directly to variation of properties characteristic of solid solution alloying of the MgB$_2$ structure.

Fig. 2 shows representative resistive transitions for film 3, in perpendicular and parallel fields up to 9 T. Also shown are the experimental definitions of $H_{c2}(T)$ and $H^*(T)$. It is immediately clear that $H^*$ and $H_{c2}$ are significantly higher in parallel than in perpendicular field. Fig. 3 shows that the $H_{c2}$ values also differ significantly from film to film, in distinct contrast to the situation in bulk samples [2-8]. For the lower resistivity films, there is curvature to the $H_{c2}(T)$ lines, but this is absent for film 1, suggestive of either a clean-to-dirty limit crossover or of variation of film properties in films 2 and 3. It is also striking that, though film 1 has the lowest $T_c$, it has the highest slope $dH_{c2}/dT$, both in parallel and perpendicular field. We measured the ratio $\eta = H_{c2}^{//}/H_{c2}^{\perp}$, which quantifies the electron mass anisotropy [10], and found that $\eta = 2.0 \pm 0.2$ for film 1, $\eta = 1.9 \pm 0.2$ for film 2 and $\eta = 1.8 \pm 0.2$ for film 3. These results are summarized in Table I. Because of the much better texture of our films, the ratios of $H_{c2}^{//}/H_{c2}^{\perp}$ are all higher than the value of 1.1 determined on imperfectly textured, hot deformed bulk MgB$_2$ by Handstein et al. [15], and are closer to the $\eta = 1.73$ for aligned particles extracted from bulk MgB$_2$ examined by de Lima et al. [16].

The resistive data in figs. 3 and 4 exhibit an interesting trend, namely that the upper critical field lines become steeper and the $T_c$ values decrease as the resistivity increases. This change of $H_{c2}(T)$ with resistivity is consistent with the dirty-limit behavior of the BCS theory [27]. This conclusion is also consistent with the fact that resistivity values of our films at $T_c$ are 2-3 orders of magnitude higher than those of the sintered bulk samples made at Ames [2,5], for which clean –limit BCS behavior and an electron mean-free path of $l \sim 60$ nm were inferred [5]. Since $l$ scales inversely with resistance, we tentatively conclude that all our films are in the dirty limit $l < \xi_0$, where $\xi_0 = (\phi_0/2\pi\mu_0 H_{c2})^{-1/2}$ is the coherence length, and $\phi_0$ is the flux quantum.

Figure 4 shows in detail the resistively determined $H^*(T)$ and $H_{c2}(T)$ lines for film 1. Also plotted are the Kramer-function extrapolations of the magnetically determined $J_c(H,T)$ from [18]. There is excellent agreement between the two $H^*(T)$ measurements. In contrast to the significant separation between $H^*(T)$ and $H_{c2}(T)$ for untextured bulk samples, where $H^*(T) \sim 0.5 H_{c2}(T)$ [2-8], here we find that $H^{*\perp}(T)$ is $\sim 0.8 H_{c2}^{\perp}(T)$ for textured films. It is now clear why $H^*(T)$ is ~0.5 $H_{c2}(T)$ for bulk untextured samples: the continuous supercurrent path is cut off at $H^*(T)^{\perp}$ for those grains aligned perpendicular to the applied field, while the measured $H_{c2}(T)$ occurs at $H_{c2}(T)^{//}$.

To address the mechanisms which determine the irreversibility line, we consider the possibility that thermal depinning occurs at $H^*(T)$, that is when the mean-squared displacements of vortex lines $u^2(T,H^*)$ become equal to $\xi^2(T)$ [10,28]. Assuming the usual temperature dependence $\xi(T) = \xi_0(1 - t^2)^{-1/2}$ with $t = T / T_c$, we calculate $H^*(T)$ using the formula for $u^2(T,H^*)$ for a uniaxial superconductor for field parallel to the c-axis [28]. We obtained the parametric expression $H^*(T) = b(t)H_{c2}(T)$, where $b(t)$ is determined by the equation $t^2 = g(b)/[\alpha^2 + g(b)]$, $g(b) = b(1 - b)^3 \ln[2 + (2b)^{-1/2}]$ and $\alpha$ quantifies the strength of vortex thermal fluctuations.

$$\alpha = \frac{4\sqrt{2}\mu_0 \pi \eta \kappa^2 \xi_0^{ab} k_B T_c}{\phi_0^2} . \quad (1)$$

Here $\kappa$ is the in-plane Ginzburg-Landau parameter, $\xi_0^{ab} = [\phi_0/2\pi\mu_0 H_{c2}^{\perp}(0)]^{1/2}$, and $\eta = H_{c2}^{//}/H_{c2}^{\perp}$ is the anisotropy parameter. For values characteristic of *untextured bulk* MgB$_2$, $T_c = 40$ K, $\kappa = 30$, $\xi_0 = 5$ nm [2,3], Eq. (1) yields $\alpha \approx 0.03$ if $\eta$ is assumed to be 2. The small value of $\alpha$ indicates that vortex fluctuations in bulk MgB$_2$ are indeed weaker than in high-$T_c$ oxides, for which $\alpha \sim 1$ [28]. The primary effect of the high resistivity of our thin films is to raise the value of $\kappa^2\xi_0^{ab} = (\lambda^{ab})^2 / \xi^{ab}$, while somewhat reducing $T_c$. Therefore, $\alpha$ should be higher in our thin films than the bulk estimate. For now we lack direct measurements of the penetration depth $\lambda^{ab}$. Consistent with the reduction in $\xi(T)$ in our alloyed films, the best fit to the experimental data, for which $H^*(T) \sim 0.8\, H_{c2}^\perp$, occurs for $\alpha = 0.08$, as shown in the inset of Fig. 4. This 20% suppression of $H^*$ below $H_{c2}$ is rather comparable to the isotropic low-$T_c$ materials such as Nb-Ti and Nb$_3$Sn for which $H^*(T) \sim 0.85\text{-}0.9\, H_{c2}$ [29].

We summarize our findings in Table I and in the schematic phase diagram of Fig. 5, for both bulk [3] and thin film MgB$_2$ samples, assuming dirty-limit extrapolations of $H_{c2}(T)$ using $H_{c2}(0) = 0.69\, T_c\, dH_{c2}/dT$ [27]. We evaluate $dH_{c2}/dT$ at fields between 3 and 9 T so as to exclude minority phases of higher $T_c$ that may be present in films 2 and 3 (this may be one cause of the $H_{c2}$ curvature evident in Fig. 3). The general picture indicated by our data is, first, that it is possible to alloy MgB$_2$ so that $H_{c2}(0)$ increases in a manner qualitatively consistent with the expression $\mu_0 H_{c2}(0) = 3110\, \rho\, \gamma\, T_c$ (all quantities being in SI units), where $\gamma$ is the electronic specific heat coefficient [27]. As noted in [18], we believe that solid solution alloying by O is the most likely source of the enhanced thin film properties that we observe. Unfortunately the approximate doubling of $H_{c2}$ is accompanied by a decrease in $T_c$ to ~30 K. Recent irradiation experiments suggest a similar increase in $H^*$ also at the expense of depressing $T_c$ [30]. A second important point is that although there is still no sign that grain boundaries are barriers to current flow, even in these very fine grain-size films [18], our results make it clear that uniaxial texturing will be needed to provide a capability significantly beyond Nb$_3$Sn, for which $H^*(4.2\text{ K})$ is ~25 T. Assuming that $H^{*\parallel}(4.2\text{ K})$ of thin film MgB$_2$ is ~$0.8 H_{c2}^\parallel$, then $H^{*\parallel}$ should exceed 31 T. However, untextured forms of MgB$_2$ would be limited to about $0.8 H_{c2}^\perp$, that is ~15 T at 4.2 K. Therefore, although the anisotropy of MgB$_2$, is much smaller than for the high temperature cuprate superconductors, it nonetheless significantly affects the magnetic field behavior and current-carrying capability of MgB$_2$. Finally we note that, though shortening the coherence length of MgB$_2$ (Table I) is the basis of improved high field performance, the ability to maintain $\xi$ at clean limit values may be very advantageous for electronic applications, especially for fabrication of superconductor / normal-metal / superconductor (SNS) and superconductor / insulator / superconductor (SIS) junctions. Understanding how to control the properties of MgB$_2$ by alloying will be very important for all applications and perhaps for understanding its superconducting mechanism.

**Acknowledgments**

The work at UW has been supported by AFOSR, DOE and the NSF though the MRSEC on Nanostructured Materials. We are grateful to Professor Robert Cava at Princeton for some of the MgB$_2$ used for film targets.

Table 1: Summary of thin film properties. Textured bulk data were obtained from [16]. The critical temperature is defined at 50% of the normal state resistance in zero field.

| Sample | $T_c$ (K) | ρ (μΩ·cm) | $\mu_0 H_{c2}^{\perp}(0)$ (T) | $\mu_0 H_{c2}^{\parallel}(0)$ (T) | η | $\xi_0^{ab}$ (nm) | $\xi_0^c$ (nm) |
|---|---|---|---|---|---|---|---|
| Film 1 | 31 | 360 | 19.5 | 39.0 | 2.0 | 4.0 | 2.0 |
| Film 2 | 32.5 | 40 | 12.7 | 24.1 | 1.9 | 5.0 | 2.6 |
| Film 3 | 37 | 38 | 12.5 | 22.5 | 1.8 | 5.0 | 2.8 |
| Textured bulk | 39 | | 6.5 | 11 | 1.7 | 7.0 | 4.1 |

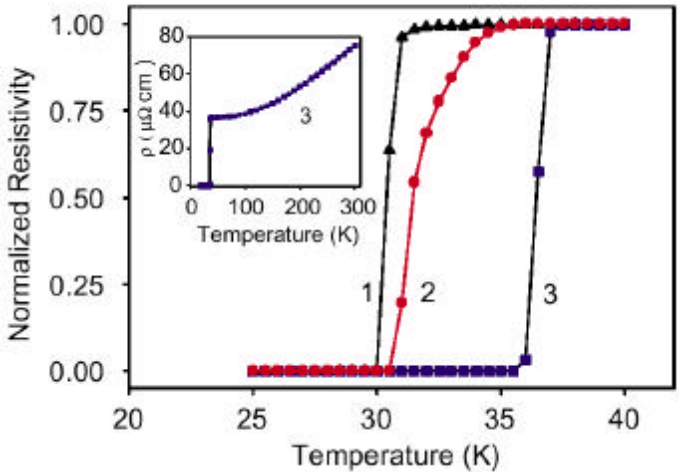

Fig.1: Zero field resistive transition of film1 (black ▲), film 2 (red ●) and film 3 (blue ■). The inset shows resistivity of film 3 up to room temperature.

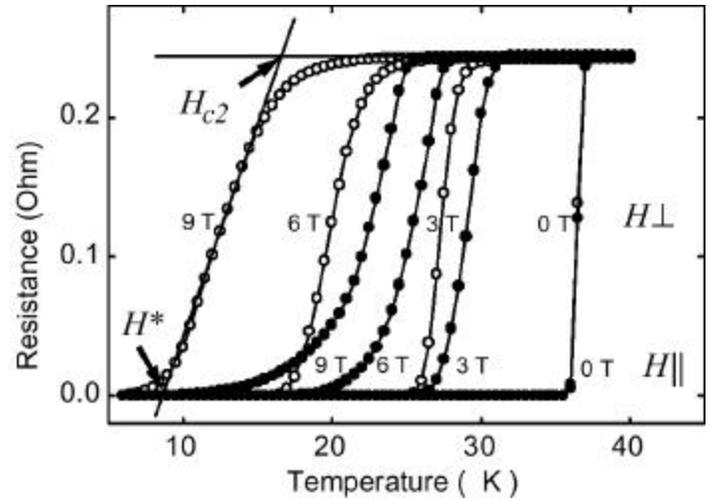

Fig. 2: Resistive transitions of film 3 in magnetic fields of (left to right) 9, 6, 3 and 0 T. The open and closed symbols are for external field applied perpendicular and parallel to the film plane, respectively. The anisotropy in upper critical field is evident in the different transition temperatures for a given applied field. The line drawn through the steepest part of the 9 T, perpendicular field data indicates how the values of $H^*$ and $H_{c2}$ were determined.

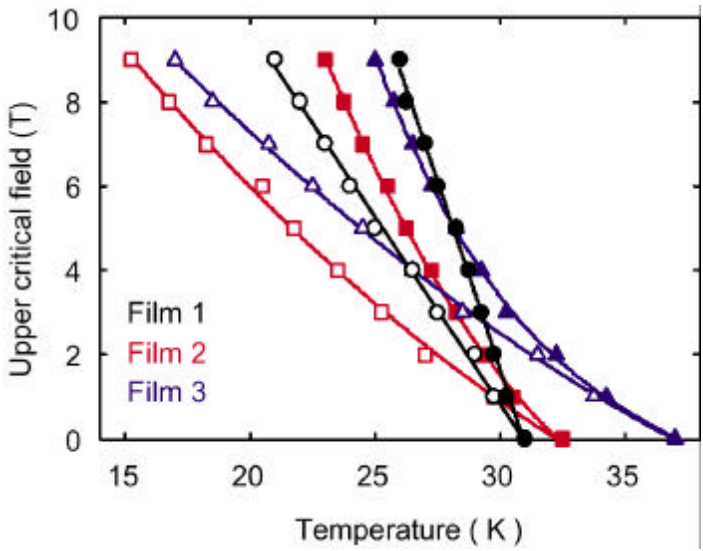

Fig. 3: The upper critical field as a function of temperature for the three films. The open and closed symbols are for field applied perpendicular and parallel to film plane, respectively. The black lines represent the best linear fit for high-resistivity film 1. Curves drawn through lower-resistivity film 2 (red) and film 3 (blue) are guides to the eye.

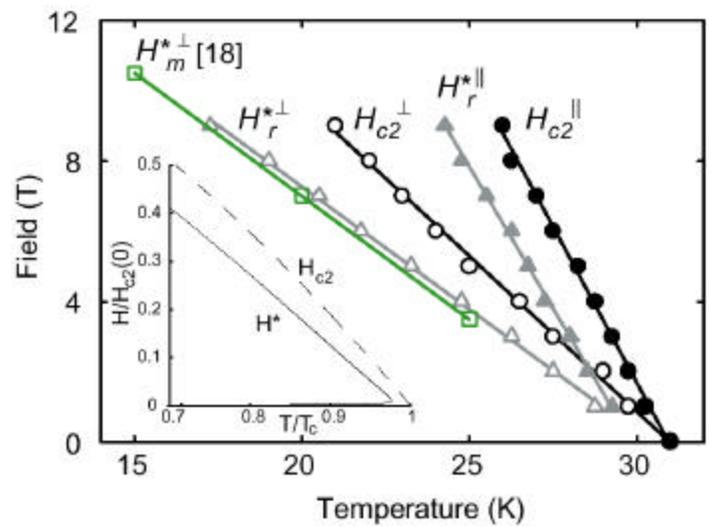

Fig. 4: The upper critical field and irreversibility lines determined by the resistive measurements for film 1. Also shown (green line) is the irreversibility line from [18] based on magnetization data. Inset: calculated thermal depinning lines for $\alpha = 0.08$ (see text), in reduced field and temperature coordinates. The dashed line indicates the perpendicular upper critical field.

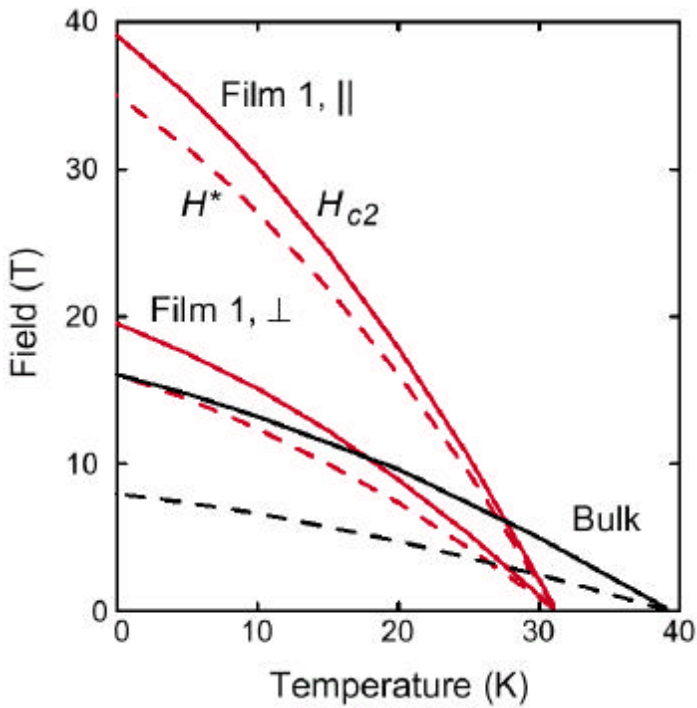

Fig. 5: Schematic representation of the field-temperature boundaries of superconductivity for bulk $MgB_2$ [3] and for film 1, assuming dirty-limit extrapolations of data near $T_c$ to low temperature are valid.